\documentclass[aps,prd,10pt,showpacs]{revtex4-2}
\usepackage{amsfonts}
\usepackage{amsmath}
\usepackage{amssymb}
\usepackage{graphicx}
\usepackage{color}

\usepackage[utf8]{inputenc}
\usepackage{epsfig}
\usepackage{bm}
\usepackage{hyperref}
\usepackage{times}

\def\be{\begin{equation}}
\def\te{\end{equation}}
\def\nn{\nonumber\\}
\def\ba{\end{eqnarray}}
\def\ea{\end{eqnarray}}
\def\bea{\begin{eqnarray}}
\def\tea{\end{eqnarray}}

\begin{document}

\title{Statistical closures from the Martin, Siggia and Rose approach to turbulence}

\author{Esteban Calzetta}
\email{calzetta@df.uba.ar}
\affiliation{Universidad de Buenos Aires, Facultad de Ciencias Exactas y Naturales, Departamento de Física, Buenos Aires, Argentina,\\
and CONICET-Universidad de Buenos Aires, Instituto de Física de Buenos Aires (IFIBA), Buenos Aires, Argentina  }

\begin{abstract}
The goal of this paper is to study the statistical closures suggested by the Martin-Siggia and Rose approach to statistical turbulence. We find that the formalism leads to a Bethe-Salpeter equation for the three point correlation of the velocity field. In the leading order approximation this equation becomes an explicit expression. We discuss under which approximations this closure reduces to that proposed in W D McComb and S R Yoffe, A formal derivation of the local energy transfer (LET) theory of homogeneous turbulence, J. Phys. A: Math. Theor. 50,  375501 (2017). This suggests ways to improve upon this closure by dropping these restrictions, resumming the perturbative expansion and/or applying renormalization group techniques.
\end{abstract}

\maketitle

\section{Introduction}
The goal of this paper is to discuss the solution of the closure problem in the field theory approach to statistical turbulence  \cite{EF10,Wyld61,Lee65,LB87}, more specifically within the Martin - Siggia and Rose (MSR) framework \cite{MSR73,Ph77,Eyink,Kamenev,JZEC,DeDGia06}. As shown by Krommes \cite{Krommes1978,Krommes1997,Krommes2002}, the MSR approach leads to a Bethe-Salpeter equation \cite{BS2,IZ,TK,Mariano} for the three point correlation of the turbulent velocity. In the simplest (one-loop) approximation, this equation reduces to an explicit formula, which under further approximations reproduces the closure advanced by McComb and Yoffe in ref. \cite{LET}.

Let us describe the closure problem in greater detail. Our starting point is the randomly driven Navier-Stokes Equation (NSE) \cite{vKH,Karman48,Chandra,MonYag71,FRIS95}

\be
P^i\left[v\right]=v^i_{,t}+v^j\partial_jv^i-\nu\bm{\Delta}v^i=f^i-\partial_ip,
\label{LIBNSE}
\te 
where 
$\nu$ is the molecular viscosity. We regard the pressure $p$ as a Legendre multiplier enforcing the constraint

\be
\partial_iv^i=0.
\te
In what follows, we shall assume the velocity fields are divergenceless and we shall omit the pressure term. The $f^i$, which also obey $\partial_if^i=0$, are the random forces, which we assume to be Gaussian with zero mean.

The problem is to compute the two-point correlation function 

\be 
C^{ij}\left(x,t;x',t'\right)=
\left\langle v^i\left({x},t\right)v^j\left({x'},t'\right)\right\rangle.
\label{2cor}
\te 
The usual approach is simply to multiply eq. (\ref{LIBNSE}) by $v^j\left({x'},t'\right)$ and take the expectation value of the whole expression \cite{vKH}. Then the right hand side may be computed from Novikov' s formula \cite{Novikov,Novikov1} (see below eq. (\ref{Nov})), but in the left hand side there is a term involving the three point function

\be 
\left\langle \left(v^j\partial_jv^i\right)\left({x},t\right)v^j\left({x'},t'\right)\right\rangle.
\label{3cor}
\te 
We realize we have not found a closed equation but just the first rung in a hierarchy, since the corresponding equation for the three point function will involve a fourth order correlation and so on. The closure problem consists in finding a way to express the three point function eq. (\ref{3cor}) as a functional of the correlation eq. (\ref{2cor}) , thus truncating the hierarchy. Our goal is to see what truncation emerges from the MSR approach, and to compare it with existing results in the literature.

A key point in the analysis of ref. \cite{LET} is to formulate the problem in such a way that the full correlation function eq. (\ref{2cor}) appears as an unknown, rather than performing a perturbative expansion for it. To replicate this within the MSR approach we must go beyond the 1-particle irreducible (1PI) effective action (EA) (see \cite{CalHu08,Ram07,Haehl18,MGKC,C25}). In this work we shall restrict ourselves to the 2-particle irreducible (2PI) EA, introduced in \cite{LutWar60,DomMar64b,DahLas62,cornwall:1974a}, see also \cite{CalHu08,Meg4,Gabor,Emil,Urko}; for a discussion of 2PI techniques in a MSR framework see \cite{Krommes1978,C09,Bode}. We leave the consideration of higher irreducible EA's for future work \cite{dch,cddn,stobol,stobol2,Berges,Meg1,Meg2}.

Concretely, our goal is to compare the closure relations obtained from the Bethe-Salpeter equation derived from MSR \cite{Krommes1978} to the closure relation advanced in ref. \cite{LET}. In \cite{LET} the problem is studied in the framework of the Fokker-Planck equation for the probability density function (pdf) of equal-time velocity fluctuations, introduced by Hopf \cite{Hopf}. A perturbation scheme is proposed where at ``zeroth order'' the pdf is assumed to be Gaussian, and then a ``first order'' closure relation is derived. Our goal is to find out what is the equivalent approximation scheme within the MSR approach. The relationship between the Hopf and MSR approaches is reviewed in \cite{C25}.

The paper is organized as follows. In the next section \ref{2PIEA} we present the 2PI MSR effective action; for a full discussion see \cite{Krommes1978,CalHu08,C09}. In section \ref{BSE} we review the derivation of the Bethe-Salpeter equation for the three point correlations \cite{Krommes1978}, and in section \ref{MCY} we derive the closure approximation which follows from a first order truncation of the Bethe-Salpeter equation. In section \ref{comparison} we compare this closure to the one proposed by McComb and Yoffe's in ref \cite{LET}. We find the necessary assumptions which are the correlate of the assumption of near Gaussianity of the Hopf pdf. Under these assumptions, both closure relations are equivalent. We conclude with some brief final remarks in section \ref{FR}.

In the Appendix we show a concrete example where the identity between the derivatives of the velocity with respect of the stirring forces and with respect to the initial velocity field may be stablished by an independent argument from the formal proof presented in section \ref{comparison}.
\section{The Martin-Siggia and Rose 2PI Effective Action}\label{2PIEA}

Let us return to the randomly driven NSE eq. (\ref{LIBNSE}) and write the noise self correlation as

\be
\left\langle f^i\left({x},t\right)f^j\left({x'},t'\right)\right\rangle=\delta\left(t-t'\right)N^{ij}\left({x-x'}\right),
\te 
where

\be
N^{ij}\left({x}\right)=\int\frac{d^3k}{\left(2\pi\right)^3}\;e^{i{kx}}\Delta^{ij}\left[k\right]N\left(k\right),
\te
$k=\left|{k}\right|$ and 

\be
\Delta^{ij}\left[k\right]=\delta^{ij}-\frac{k^ik^j}{k^2}.
\te

We shall assume the NSE (\ref{LIBNSE}) are the continuum limit of a discrete time evolution

\be
v_{k+1}^i\left(x\right)-v_{k}^i\left(x\right)=dt\int d^dx'\;\Delta^i_{i'}\left(x,x'\right)\left\{\nu \bm{\Delta}v_{k}^{i'}\left(x'\right)-v_{k}^j\left(x'\right)v_{k,j}^{i'}\left(x'\right)+f_{k}^{i'}\left(x'\right)\right\},
\label{DNSE}
\te
where $v_k^i\left(x^j\right)=v^i\left(x^j,t_k\right)$, $t_k=kdt$; $\Delta$ is the projector on divergenceless fields

\be
\Delta^i_{i'}\left(x,x'\right)=\int\frac{d^3k}{\left(2\pi\right)^3}\;e^{i{k\left(x-x'\right)}}\Delta^{ij}\left[k\right].
\te
We also assume that the NSE admit only one solution for given initial conditions and realization of the driving forces, which moreover are regarded as independent of each other. This requires breaking random Galilean invariance \cite{C25}. For example, we may assume that the total linear momentum of the fluid vanishes, which in turn requires the random sources to average to zero over large scales. We shall assume the initial conditions have been set far enough in the past that any memory of them has been erased, and that a statistically homogeneous and isotropic configuration has been achieved.

We introduce a generating functional for the velocity field

\bea
e^{iW\left[j_V\right]}&=&\int Df F\left[f\right]e^{i\int d^dxdt\;j_{Vk}v^k\left[t;f\right]}\nn
&=&\int DfDv F\left[f\right]\delta\left(v^k-v^k\left[t;f\right]\right)e^{i\int d^dxdt\;j_{Vk}v^k}.
\tea 
Here, $F$ is the noise probability density function

\be
F\left[f\right]= e^{-\frac12\int d^3xd^3x'dt\;f^i\left(x,t\right)N^{-1}_{ij}\left(x-x'\right)f^j\left(x',t\right)}.
\te
We assume that normalization constants are already included in the integration measure.  $v^k\left[t;f\right]$ is the unique solution to the NSE for the given random driving. We may use the identity (cfr. eq. (\ref{LIBNSE}))

\be
\delta\left(v^i\left(t\right)-v^i\left[t;f^i\right]\right)=\left({\rm{Det}}\;\frac{\delta P^i\left[v\right]}{\delta v^j}\right)\delta\left(P^i-f^i\right).
\te
Under the discretization rules above the determinant has value $1$ \cite{Zin93}. We exponentiate the delta function adding a divergenceless auxiliary field $A_i$ and integrate over the driving forces to get

\be
e^{iW\left[j_{Vk}\right]}=\int DADv\; e^{-\frac12\int d^3xd^3x'dt\;A_i\left(x,t\right)N^{ij}\left(x-x'\right)A_j\left(x',t\right)+i\int d^3xdt\;\left[A_i\left(x,t\right)P^i\left[v\right]+j_{Vk}v^k\right]}.
\te 
To obtain a more symmetric form we also add a source for the auxiliary field $A_i$

\be
e^{iW\left[j_{V},j_A\right]}=\int DADv e^{-\frac12\int d^3xd^3x'dt\;A_i\left(x,t\right)N^{ij}\left(x-x'\right)A_j\left(x',t\right)+i\int d^3xdt\;\left[A_iP^i+j_{Vk}v^k+j_A^lA_l\right]}.
\te 
Observe that

\be
W\left[0,j_A\right]\equiv 0,
\te 
which shows that the expectation value of any product of auxiliary fields vanishes on shell, namely when the external sources vanish.

We are interested in a statistically homogeneous and isotropic situation where the mean velocity $\bar v^i$ and mean auxiliary field $\bar A_j$ vanish and the relevant information is carried by the correlation functions $\left\langle v^i\left(x,t\right)v^j\left(x',t'\right)\right\rangle$ and $\left\langle v^i\left(x,t\right)A_j\left(x',t'\right)\right\rangle$ (since $\left\langle A_i\left(x,t\right)A_j\left(x',t'\right)\right\rangle=0$). To obtain self-consistent equations for the correlations we further  introduce non local sources directly coupled to the products of fields

\bea
&&e^{iW\left[j_{V},j_A,J_{VV},J_{VA},J_{AV},J_{AA}\right]}=\int DADv\; e^{iS\left[A,v\right]+i\int d^dxdt\;\left[j_{Vk}v^k+j_A^lA_l\right]}\nn
&&\exp\left\{\left(i/2\right)\int d^dxdtd^dx'dt'\left[J_{VVjk}\left(x,t;x't'\right)v^j\left(x,t\right)v^k\left(x',t'\right)\right.\right.\nn
&+&\left.\left.J_{VAj}^k\left(x,t;x't'\right)v^j\left(x,t\right)A_k\left(x',t'\right)+J_{AVk}^jA_j\left(x,t\right)v^k\left(x',t'\right)+J_{AA}^{jk}A_j\left(x,t\right)A_k\left(x',t'\right)\right]\right\},
\tea 
where $S$ is the Martin-Siggia-Rose (MSR) action 

\be
S=S_q+S_c+iS_N,
\te 
where 

\bea
S_q&=&\int d^dxdt\;A_i\left(x,t\right)\left[\partial_t-\nu\bm{\Delta}\right]v^i\left(x,t\right)\nn
S_c&=&-\frac12\int d^dxdt\;\left[A_{i,j}+A_{j,i}\right]\left(x,t\right)v^i\left(x,t\right)v^j\left(x,t\right)\nn
S_N&=&\frac 12\int d^3xd^3x'dt\;A_i\left(x,t\right)N^{ij}\left(x-x'\right)A_j\left(x',t\right).
\tea
The first derivatives of the generating functional define the mean fields and correlations

\bea
\frac{\delta W}{\delta j_{Vj}}&=&\bar v^j=\left\langle v^j\right\rangle\nn
\frac{\delta W}{\delta j_{A}^k}&=&\bar A_k=\left\langle A_k\right\rangle\nn
\frac{\delta W}{\delta j_{VVjk}}&=&\frac12\left\langle v^jv^k\right\rangle=\frac12\left[ \bar v^j\bar v^k+\left\langle \left\langle v^jv^k\right\rangle\right\rangle\right]\nn
\frac{\delta W}{\delta j_{VAj}^k}&=&\frac12\left\langle v^jA_k\right\rangle=\frac12\left[ \bar v^j\bar A_k+\left\langle \left\langle v^jA_k\right\rangle\right\rangle\right]\nn
\frac{\delta W}{\delta j_{AVk}^j}&=&\frac12\left\langle A_jv^k\right\rangle=\frac12\left[ \bar A_j\bar v^k+\left\langle \left\langle A_jv^k\right\rangle\right\rangle\right]\nn
\frac{\delta W}{\delta j_{AA}^{jk}}&=&\frac12\left\langle A_jA_k\right\rangle=\frac12\left[ \bar A_j\bar A_k+\left\langle \left\langle A_jA_k\right\rangle\right\rangle\right],
\tea
where a double bracket such as $\left\langle \left\langle v^jv^k\right\rangle\right\rangle$ means an irreducible correlation

\be
\left\langle \left\langle v^jv^k\right\rangle\right\rangle=\left\langle \left(v^j-\bar v^j\right)\left(v^k-\bar v^k\right)\right\rangle.
\te
When it is clear that the mean fields vanish we shall drop the double brackets.

The 2-particle irreducible effective action (2PIEA) is the full Legendre transform

\bea
&&\Gamma\left[\bar v,\bar A,\left\langle \left\langle vv\right\rangle\right\rangle,\left\langle \left\langle vA\right\rangle\right\rangle,\left\langle \left\langle Av\right\rangle\right\rangle,\left\langle \left\langle AA\right\rangle\right\rangle\right]=W\left[j_{V},j_A,J_{VV},J_{VA},J_{AV},J_{AA}\right]\nn
&-&\int d^dxdt \left[j_{Vj}\bar v^j+j_A^k\bar A_k\right]\nn
&-&\frac12\int d^dxdt d^dx'dt' \left[J_{VVjk}\left[ \bar v^j\bar v^k+\left\langle \left\langle v^jv^k\right\rangle\right\rangle\right]+J_{VAj}^k\left[ \bar v^j\bar A_k+\left\langle \left\langle v^jA_k\right\rangle\right\rangle\right]\right.\nn
&+&\left.J_{AVk}^j\left[ \bar A_j\bar v^k+\left\langle \left\langle A_jv^k\right\rangle\right\rangle\right]+J_{AA}^{jk}\left[ \bar A_j\bar A_k+\left\langle \left\langle A_jA_k\right\rangle\right\rangle\right]\right],
\tea 
whereby the mean fields obey the equations of motion

\bea
\frac{\delta\Gamma}{\delta\bar v^j}&=&-j_{Vj}-\int d^dx'dt'\left[J_{VVjk}\bar v^k+\frac12\left(J_{VAj}^k+\left.J_{AVj}^k\right)\bar A_k\right]\right]\nn
\frac{\delta\Gamma}{\delta\bar A_j}&=&-j_{Aj}-\int d^dx'dt' \left[\frac12\left(J_{VAk}^j+\left.J_{AVk}^j\right)\bar v^k+J_{AA}^{jk}\bar A_k\right]\right]\nn
\frac{\delta\Gamma}{\delta\left\langle \left\langle v^jv^k\right\rangle\right\rangle}&=&-\frac12J_{VVjk}\nn
\frac{\delta\Gamma}{\delta\left\langle \left\langle v^jA_k\right\rangle\right\rangle}&=&-\frac12J_{VAj}^k\nn
\frac{\delta\Gamma}{\delta\left\langle \left\langle A_jv^k\right\rangle\right\rangle}&=&-\frac12J_{AVk}^j\nn
\frac{\delta\Gamma}{\delta\left\langle \left\langle A^jA^k\right\rangle\right\rangle}&=&-\frac12J_{AA}^{jk}.
\tea 
The 2PIEA has the structure \cite{CalHu08}

\be
\Gamma=S\left[\bar v,\bar A\right]+\Gamma_1+\Gamma_2
\te
where 

\bea
&&\Gamma_1=\frac12{\rm{tr}}\;\left(\begin{array}{cc}\frac{\partial^2S}{\partial \bar v^j\partial\bar v^k}&\frac{\partial^2S}{\partial \bar v^j\partial\bar A_k}\\\frac{\partial^2S}{\partial \bar A_j\partial\bar v^k}&\frac{\partial^2S}{\partial \bar A_j\partial\bar A_k}\end{array}\right)\left(\begin{array}{cc}\left\langle \left\langle v^kv^l\right\rangle\right\rangle&\left\langle \left\langle v^kA_l\right\rangle\right\rangle\\\left\langle \left\langle A_kv^l\right\rangle\right\rangle&\left\langle \left\langle A_kA_l\right\rangle\right\rangle\end{array}\right)\nn
&-&\frac i2\ln\;{\rm{det}}\;\left(\begin{array}{cc}\left\langle \left\langle v^kv^l\right\rangle\right\rangle&\left\langle \left\langle v^kA_l\right\rangle\right\rangle\\\left\langle \left\langle A_kv^l\right\rangle\right\rangle&\left\langle \left\langle A_kA_l\right\rangle\right\rangle\end{array}\right).
\tea
$\Gamma_2$ is the sum of all 2 particle irreducible (2PI) vacuum bubbles built from the cubic vertex extracted from $S_c$ and irreducible correlations in the internal legs. Note that $\Gamma_2$ is independent of the mean fields, because the interaction is cubic.

Observe that a 2PI graph must be necessarily connected and that the simplest vacuum bubble built from cubic vertices contains two loops. In general, consider a vacuum bubble with $V$ vertices, $L$ loops and $I_{vv}$, $I_{vA}$ and $I_{AA}$ internal lines, where $I_{xy}$ counts lines containing an $\left\langle \left\langle xy\right\rangle\right\rangle$ correlation. Then we have

\bea
2V&=&2I_{vv}+I_{vA}\nn
V&=&I_{vA}+2I_{AA}\nn
V&=&I_{vv}+I_{vA}+I_{AA}-L+1,
\tea
whereby

\bea
V&=&2\left(L-1\right)\nn
I_{vA}&=&2\left(L-1-I_{AA}\right)\nn
I_{vv}&=&L-1+I_{AA}.
\tea
This shows there are no 2PI vacuum bubbles with just one loop.

\section{Bethe-Salpeter equation for three-point correlations}\label{BSE}

Let us write again the equation of motion derived from the variation of the 2PIEA with respect to $\bar A_m$ in the case where all non local sources are set to zero,

\be
\bar v^i_{,t}+\left(\bar v^j\partial_j\bar v^i\right)_{\perp}-\nu\bm{\Delta}\bar v^i+i\int d^dyds\;N^{ij}\left(x,t;y,s\right)\bar A_j\left(y,s\right)+\left\langle \left\langle \left(v^j\left(x,t\right)\partial_jv^i\left(x,t\right)\right)_{\perp}\right\rangle\right\rangle=-j_{A}^i,
\label{master}
\te
where for any vector field $V^i$

\be
V^i_{\perp}\left(x,t\right)=\int d^dx'\;\Delta^i_{i'}\left(x,x'\right)V^{i'}\left(x',t\right).
\te
The correlations themselves are given by

\be
\left(\begin{array}{cc}\frac{\partial^2S}{\partial \bar v^j\partial\bar v^k}&\frac{\partial^2S}{\partial \bar v^j\partial\bar A_k}\\\frac{\partial^2S}{\partial \bar A_j\partial\bar v^k}&\frac{\partial^2S}{\partial \bar A_j\partial\bar A_k}\end{array}\right)-i\left(\begin{array}{cc}\left\langle \left\langle v^kv^j\right\rangle\right\rangle&\left\langle \left\langle v^kA_j\right\rangle\right\rangle\\\left\langle \left\langle A_kv^j\right\rangle\right\rangle&\left\langle \left\langle A_kA_j\right\rangle\right\rangle\end{array}\right)^{-1}+2\left(\begin{array}{cc}\frac{\partial\Gamma_2}{\partial \left\langle \left\langle v^jv^k\right\rangle\right\rangle}&\frac{\partial\Gamma_2}{\partial \left\langle \left\langle v^jA_k\right\rangle\right\rangle}\\\frac{\partial\Gamma_2}{\partial \left\langle \left\langle A_jv^k\right\rangle\right\rangle}&\frac{\partial\Gamma_2}{\partial \left\langle \left\langle A_jA_k\right\rangle\right\rangle}\end{array}\right)=0.
\label{master2}
\te
We may use equations (\ref{master}) and (\ref{master2}) to find the correlations on shell, where the mean fields and also $\left\langle A_jA_k\right\rangle=0$. However, it is best to derive equations where the three point correlations appear explicitly. To this end, we take derivatives of equations  (\ref{master}) and (\ref{master2}) with respect to the local sources $j_v$ and $j_A$ in turn, and then set these sources to zero (thereby turning off the mean fields too). Recall the derivative rules

\bea
\frac{\delta\left\langle X\right\rangle}{\delta j_{Vk}}&=&i\left\langle \left\langle Xv^k\right\rangle\right\rangle\nn
\frac{\delta\left\langle X\right\rangle}{\delta j_{A}^k}&=&i\left\langle \left\langle XA_k\right\rangle\right\rangle.
\tea
From the derivative of eq. (\ref{master})with respect to $j_v$ we get

\bea
&&\left\langle v^i_{,t}\left(x,t\right)v^k\left(x',t'\right)\right\rangle-\nu\left(\bm{\Delta}v^i\left(x,t\right)v^k\left(x',t'\right)\right\rangle+\left\langle \left(v^j\partial_jv^i\right)_{\perp}\left(x,t\right)v^k\left(x',t'\right)\right\rangle\nn
&=&-i\int d^dy\;N^{ij}\left(x-y\right)\left\langle v^k\left(x',t'\right) A_j\left(y,t\right)\right\rangle.
\label{NSE1}
\tea
The derivative of eq. (\ref{master})with respect to $j_A$ yields

\be
\left\langle v^i_{,t}\left(x,t\right)A_k\left(x',t'\right)\right\rangle-\nu\left(\bm{\Delta}v^i\left(x,t\right)A_k\left(x',t'\right)\right\rangle+\left\langle \left(v^j\partial_jv^i\right)_{\perp}\left(x,t\right)A_k\left(x',t'\right)\right\rangle=i\Delta^i_k\left(x-x'\right)\delta\left(t-t'\right).
\label{NSE2}
\te
We finally must face the closure problem, that is, to express the three point functions in equations (\ref{NSE1}) and (\ref{NSE2}) in terms of two point correlations. Let us introduce the matrix of field products

\be
g^{jk}=\left(\begin{array}{cc}v^kv^j& v^kA_j\\A_kv^j& A_kA_j\end{array}\right),
\te 
so we may write eq. (\ref{master2}) as

\be
\left(\begin{array}{cc}\frac{\partial^2S}{\partial \bar v^j\partial\bar v^k}&\frac{\partial^2S}{\partial \bar v^j\partial\bar A_k}\\\frac{\partial^2S}{\partial \bar A_j\partial\bar v^k}&\frac{\partial^2S}{\partial \bar A_j\partial\bar A_k}\end{array}\right)-i\left\langle g^{jk}\right\rangle^{-1}+2\frac{\partial\Gamma_2}{\partial \left\langle g^{jk}\right\rangle}=0.
\label{master3b}
\te
Now the derivative with respect to $j_V$ yields

\be
\left(\begin{array}{cc}\frac{\partial^3S}{\partial \bar v^j\partial\bar v^k\partial\bar A_l}\left\langle A_lv^m\right\rangle&\frac{\partial^3S}{\partial \bar v^j\partial\bar A_k\partial v^l}\left\langle v^lv^m\right\rangle\\\frac{\partial^2S}{\partial \bar A_j\partial\bar v^k\partial\bar v^l}\left\langle v^lv^m\right\rangle&0\end{array}\right)+i\left\langle g^{jl}\right\rangle^{-1}\left\langle g^{ln}v^m\right\rangle\left\langle g^{nk}\right\rangle^{-1}+2\frac{\delta^2\Gamma_2}{\delta \left\langle g^{jk}\right\rangle\delta\left\langle g^{ln}\right\rangle}\left\langle g^{ln}v^m\right\rangle=0.
\label{master3}
\te
which may be cast as a Bethe-Salpeter equation

\small{
\bea
&&\left( \begin{array}{c}\left\langle v^lv^nv^m\right\rangle \\ \left\langle v^lA^nv^m\right\rangle \\ \left\langle A^lv^nv^m\right\rangle \\ \left\langle A^lA^nv^m\right\rangle\end{array}\right) =i\left( \begin{array}{cccc}\left\langle v^lv^j\right\rangle \left\langle v^kv^n\right\rangle & \left\langle v^lv^j\right\rangle \left\langle A^kv^n\right\rangle & \left\langle v^lA^j\right\rangle \left\langle v^kv^n\right\rangle & \left\langle v^lA^j\right\rangle \left\langle A^kv^n\right\rangle \\
\left\langle v^lv^j\right\rangle \left\langle v^kA^n\right\rangle & 0 & \left\langle v^lA^j\right\rangle \left\langle v^kA^n\right\rangle & 0\\
\left\langle A^lv^j\right\rangle \left\langle v^kv^n\right\rangle & \left\langle A^lv^j\right\rangle \left\langle A^kv^n\right\rangle & 0 & 0 \\
\left\langle A^lv^j\right\rangle \left\langle v^kA^n\right\rangle & 0 & 0 & 0\end{array}\right) \nn
&&\left\lbrace \left( \begin{array}{c}\frac{\partial^3S}{\partial \bar v^j\partial\bar v^k\partial\bar A_p}\left\langle A_pv^m\right\rangle\\
\frac{\partial^3S}{\partial \bar v^j\partial\bar A_k\partial v^p}\left\langle v^pv^m\right\rangle\\
\frac{\partial^2S}{\partial \bar A_j\partial\bar v^k\partial\bar v^p}\left\langle v^pv^m\right\rangle\\
0\end{array}\right) +2\left( \begin{array}{cccc}
\frac{\delta^2\Gamma_2}{\delta \left\langle v^jv^k\right\rangle\delta\left\langle v^pv^q\right\rangle}&
\frac{\delta^2\Gamma_2}{\delta \left\langle v^jv^k\right\rangle\delta\left\langle v^pA^q\right\rangle}&
\frac{\delta^2\Gamma_2}{\delta \left\langle v^jv^k\right\rangle\delta\left\langle A^pv^q\right\rangle}&
\frac{\delta^2\Gamma_2}{\delta \left\langle v^jv^k\right\rangle\delta\left\langle A^pA^q\right\rangle}\\
\frac{\delta^2\Gamma_2}{\delta \left\langle v^jA^k\right\rangle\delta\left\langle v^pv^q\right\rangle}&
\frac{\delta^2\Gamma_2}{\delta \left\langle v^jA^k\right\rangle\delta\left\langle v^pA^q\right\rangle}&
\frac{\delta^2\Gamma_2}{\delta \left\langle v^jA^k\right\rangle\delta\left\langle A^pv^q\right\rangle}&
\frac{\delta^2\Gamma_2}{\delta \left\langle v^jA^k\right\rangle\delta\left\langle A^pA^q\right\rangle}\\
\frac{\delta^2\Gamma_2}{\delta \left\langle A^jv^k\right\rangle\delta\left\langle v^pv^q\right\rangle}&
\frac{\delta^2\Gamma_2}{\delta \left\langle A^jv^k\right\rangle\delta\left\langle v^pA^q\right\rangle}&
\frac{\delta^2\Gamma_2}{\delta \left\langle A^jv^k\right\rangle\delta\left\langle A^pv^q\right\rangle}&
\frac{\delta^2\Gamma_2}{\delta \left\langle A^jv^k\right\rangle\delta\left\langle A^pA^q\right\rangle}\\
\frac{\delta^2\Gamma_2}{\delta \left\langle A^jA^k\right\rangle\delta\left\langle v^pv^q\right\rangle}&
\frac{\delta^2\Gamma_2}{\delta \left\langle A^jA^k\right\rangle\delta\left\langle v^pA^q\right\rangle}&
\frac{\delta^2\Gamma_2}{\delta \left\langle A^jA^k\right\rangle\delta\left\langle A^pv^q\right\rangle}&
\frac{\delta^2\Gamma_2}{\delta \left\langle A^jA^k\right\rangle\delta\left\langle A^pA^q\right\rangle}\end{array}\right) 
\left( \begin{array}{c}\left\langle v^pv^qv^m\right\rangle \\ \left\langle v^pA^qv^m\right\rangle \\ \left\langle A^pv^qv^m\right\rangle \\ \left\langle A^pA^qv^m\right\rangle\end{array}\right)\right\rbrace \nn
\label{BSv}
\tea }

Similarly, the derivative with respect to $j_A$ yields

\bea
&&\left( \begin{array}{c}\left\langle v^lv^nA_m\right\rangle \\ \left\langle v^lA^nA_m\right\rangle \\ \left\langle A^lv^nA_m\right\rangle \end{array}\right) =i\left( \begin{array}{ccc}\left\langle v^lv^j\right\rangle \left\langle v^kv^n\right\rangle & \left\langle v^lv^j\right\rangle \left\langle A^kv^n\right\rangle & \left\langle v^lA^j\right\rangle \left\langle v^kv^n\right\rangle  \\
\left\langle v^lv^j\right\rangle \left\langle v^kA^n\right\rangle & 0 & \left\langle v^lA^j\right\rangle \left\langle v^kA^n\right\rangle \\
\left\langle A^lv^j\right\rangle \left\langle v^kv^n\right\rangle & \left\langle A^lv^j\right\rangle \left\langle A^kv^n\right\rangle & 0 \end{array}\right) \nn
&&\left\lbrace \left( \begin{array}{c}0\\
\frac{\partial^3S}{\partial \bar v^j\partial\bar A_k\partial v^p}\left\langle v^pA_m\right\rangle\\
\frac{\partial^2S}{\partial \bar A_j\partial\bar v^k\partial\bar v^p}\left\langle v^pA_m\right\rangle\end{array}\right) +2\left( \begin{array}{ccc}
\frac{\delta^2\Gamma_2}{\delta \left\langle v^jv^k\right\rangle\delta\left\langle v^pv^q\right\rangle}&
\frac{\delta^2\Gamma_2}{\delta \left\langle v^jv^k\right\rangle\delta\left\langle v^pA^q\right\rangle}&
\frac{\delta^2\Gamma_2}{\delta \left\langle v^jv^k\right\rangle\delta\left\langle A^pv^q\right\rangle}\\
\frac{\delta^2\Gamma_2}{\delta \left\langle v^jA^k\right\rangle\delta\left\langle v^pv^q\right\rangle}&
\frac{\delta^2\Gamma_2}{\delta \left\langle v^jA^k\right\rangle\delta\left\langle v^pA^q\right\rangle}&
\frac{\delta^2\Gamma_2}{\delta \left\langle v^jA^k\right\rangle\delta\left\langle A^pv^q\right\rangle}\\
\frac{\delta^2\Gamma_2}{\delta \left\langle A^jv^k\right\rangle\delta\left\langle v^pv^q\right\rangle}&
\frac{\delta^2\Gamma_2}{\delta \left\langle A^jv^k\right\rangle\delta\left\langle v^pA^q\right\rangle}&
\frac{\delta^2\Gamma_2}{\delta \left\langle A^jv^k\right\rangle\delta\left\langle A^pv^q\right\rangle}\end{array}\right) 
\left( \begin{array}{c}\left\langle v^pv^qA_m\right\rangle \\ \left\langle v^pA^qA_m\right\rangle \\ \left\langle A^pv^qA_m\right\rangle\end{array}\right)\right\rbrace \nn
\label{BSA}
\tea

\section{Lowest order MSR closure}\label{MCY}

In the simplest approximation where we neglect $\Gamma_2$ eq. (\ref{BSv}) yields

\bea
&&\left\langle v^l\left(x,t\right)v^n\left(x',t'\right)v^m\left(x'',t''\right)\right\rangle\nn
&=&i\int d^dyds\left\{\left\langle  v^l\left(x,t\right) v^j\left(y,s\right)\right\rangle\left\langle  v^n\left(x',t'\right) v^k\left(y,s\right)\right\rangle\left\langle v^m\left(x'',t''\right)\left[A_{j,k}+A_{k,j}\right]\left(y,s\right)\right\rangle\right.\nn
&+&\left\langle  v^l\left(x,t\right) v^j\left(y,s\right)\right\rangle\left\langle  v^m\left(x'',t''\right) v^k\left(y,s\right)\right\rangle\left\langle v^n\left(x',t'\right)\left[A_{j,k}+A_{k,j}\right]\left(y,s\right)\right\rangle\nn
&+&\left.\left\langle  v^n\left(x',t'\right) v^j\left(y,s\right)\right\rangle\left\langle  v^m\left(x'',t''\right) v^k\left(y,s\right)\right\rangle\left\langle v^l\left(x,t\right)\left[A_{j,k}+A_{k,j}\right]\left(y,s\right)\right\rangle\right\},
\tea
and similarly

\bea
&&\left\langle v^l\left(x,t\right)v^n\left(x',t'\right)A_m\left(x'',t''\right)\right\rangle\nn
&=&i\int d^dyds\left\{\left\langle  v^l\left(x,t\right) v^j\left(y,s\right)\right\rangle\left\langle  A_m\left(x'',t''\right) v^k\left(y,s\right)\right\rangle\left\langle v^n\left(x',t'\right)\left[A_{j,k}+A_{k,j}\right]\left(y,s\right)\right\rangle\right.\nn
&+&\left.\left\langle  v^n\left(x',t'\right) v^j\left(y,s\right)\right\rangle\left\langle  A_m\left(x'',t''\right) v^k\left(y,s\right)\right\rangle\left\langle v^l\left(x,t\right)\left[A_{j,k}+A_{k,j}\right]\left(y,s\right)\right\rangle\right\}.
\tea
Let us work out in detail the three point function

\bea
&&\left\langle v^l\left(x,t\right)v^n\left(x,t\right)v^m\left(x',t'\right)\right\rangle\nn
&=&i\int d^dyds\left\{\left\langle  v^l\left(x,t\right) v^j\left(y,s\right)\right\rangle\left\langle  v^n\left(x,t\right) v^k\left(y,s\right)\right\rangle\left\langle v^m\left(x',t'\right)\left[A_{j,k}+A_{k,j}\right]\left(y,s\right)\right\rangle\right.\nn
&+&\left\langle  v^l\left(x,t\right) v^j\left(y,s\right)\right\rangle\left\langle  v^m\left(x',t'\right) v^k\left(y,s\right)\right\rangle\left\langle v^n\left(x,t\right)\left[A_{j,k}+A_{k,j}\right]\left(y,s\right)\right\rangle\nn
&+&\left.\left\langle  v^n\left(x,t\right) v^j\left(y,s\right)\right\rangle\left\langle  v^m\left(x',t'\right) v^k\left(y,s\right)\right\rangle\left\langle v^l\left(x,t\right)\left[A_{j,k}+A_{k,j}\right]\left(y,s\right)\right\rangle\right\}.
\tea
We introduce the Fourier transforms

\bea
\left\langle  v^l\left(x,t\right) v^j\left(y,s\right)\right\rangle&=&\int\frac{d^dq}{\left(2\pi\right)^d}e^{iq\left(x-y\right)}\Delta^{lj}\left[q\right]C\left[q,t,s\right]\nn
\left\langle  v^l\left(x,t\right) A_j\left(y,s\right)\right\rangle&=&i\int\frac{d^dq}{\left(2\pi\right)^d}e^{iq\left(x-y\right)}\Delta^l_j\left[q\right]G\left[q,t,s\right].
\label{ge2}
\tea
Then

\bea
&&\left\langle v^l\left(x,t\right)v^n\left(x,t\right)v^m\left(x',t'\right)\right\rangle\nn
&=&i\int\frac{d^dp}{\left(2\pi\right)^3}\frac{d^dq}{\left(2\pi\right)^3}\frac{d^dq'}{\left(2\pi\right)^3}\;e^{ip\left(x-x'\right)}\left(2\pi\right)^3\delta\left(-p+q+q'\right)\nn
&&\int ds\;C\left[q',t,s\right]\left\{-A^{lnm}\left[q',q,p\right]C\left[q,t,s\right]G\left[p,t',s\right]+B^{lnm}\left[q',q,p\right]C\left[p,t',s\right]G\left[q,t,s\right]\right\},
\tea
where

\bea
A^{lnm}\left[q',q,p\right]&=&\Delta^{lj}\left[q'\right] \Delta^{nk}\left[q\right] \left(p_j\Delta^m_k\left[p\right]+p_k\Delta^m_j\left[p\right]\right)\nn
B^{lnm}\left[q',q,p\right]&=&\Delta^{mk}\left[p\right]\left[\Delta^{lj}\left[q'\right]\left(q_j\Delta^n_k\left[q\right]+q_k\Delta^n_j\left[q\right]\right)+\left(l\leftrightarrow n\right)\right].
\tea
It is easily seen that

\bea
p_lA^{lnm}\left[q',q,p\right]&=&\Delta^{mn}\left[p\right]A\left[q',q,p\right]\nn
p_lB^{lnm}\left[q',q,p\right]&=&\Delta^{mn}\left[p\right]B\left[q',q,p\right],
\tea 
where

\be
B\left[q',q,p\right]=\frac12p_lB^ln_{\;\;n}\left[q',q,p\right]=\frac12\left[p_l\Delta^m_k\left[p\right]+p_k\Delta^m_l\left[p\right]\right]\Delta^{lj}\left[q'\right]\left[q_j\Delta^k_m\left[q\right]+q^k\Delta_{mj}\left[q\right]\right]
\te
and 

\be
A\left[q',q,p\right]=\frac12p_lA^ln_{\;\;n}\left[q',q,p\right]=B\left[q',q,p\right]-\frac14\left[p_l\Delta_{mj}\left[p\right]+p_k\Delta_{ml}\left[p\right]\right]\Delta^{lj}\left[q'\right]\Delta^m_j\left[q\right]\left(q^k-q'^k\right).
\te 
Since $A-B$ is antisymmetric in $\left(q,q'\right)$, we may discard this term and write

\bea
&&\left\langle v^l\left(x,t\right)v^n_{,l}\left(x,t\right)v^m\left(x',t'\right)\right\rangle\nn
&=&\int\frac{d^dp}{\left(2\pi\right)^3}\frac{d^dq}{\left(2\pi\right)^3}\frac{d^dq'}{\left(2\pi\right)^3}\;e^{ip\left(x-x'\right)}\left(2\pi\right)^3\delta\left(-p+q+q'\right)\Delta^{mn}\left[p\right]B\left[q',q,p\right]\nn
&&\int ds\;C\left[q',t,s\right]\left\{C\left[q,t,s\right]G\left[p,t',s\right]-C\left[p,t',s\right]G\left[q,t,s\right]\right\},
\label{MSRC}
\tea
which is the lowest order closure from MSR. 

\section{MSR and LET closures, a comparison}\label{comparison}

The closure given in eq. (4.55) of ref. \cite{LET} is identical to the one given in eq. (\ref{MSRC}), except that, instead of the function $G\left[p,t,s\right]$ defined in eq. (\ref{ge2}), a new function $R\left[p,t',s\right]$ appears. The new function comes from the Fourier transform

\be 
\left\langle \frac{\partial v^j\left(x,t\right)}{\partial v^k\left(y,t'\right)}\right\rangle=R^j_k[x,t;y,t']=
\int\frac{d^dp}{\left(2\pi\right)^d}e^{ip\left(x-y\right)}\Delta^j_k\left[p\right]R\left[p,t,s\right].
\label{erre}
\te
Moreover, the new function $R$ only needs to be known to zeroeth order in the LET perturbative scheme. To this order equal time velocity fluctuations are assumed to be Gaussian, and then from Novikov's formula \cite{Novikov,Novikov1} we obtain the so-called ``fluctuation-response relation'', eq. (4.68) in ref. \cite{LET},

\be
C[p,t,s]=R[p,t,s]C[p,s,s]\;{\rm{for}}\;t>s.
\label{frr}
\te 
We emphasize that eq. (\ref{frr}) only holds when the pdf of equal time velocity fluctuations is Gaussian, or under the even more restrictive assumption that the evolution is linear. We shall assume the former. 

To prove the equivalence of the closure eq.(\ref{MSRC}) to the LET closure we must show that actually the two functions $G[p,t,t']$ and $R[p,t,t']$ are the same for all times $t>t'$, and then provide the supplementary assumptions under which we recover a MSR ``fluctuation-response relation''

\be
C[p,t,s]=G[p,t,s]C[p,s,s]\;{\rm{for}}\;t>s
\label{MSRfrr}
\te 
These supplementary assumptions are the MSR counterpart of the near Gaussianity assumed in LET.

To show that $G[p,t,t']=R[p,t,t']$ when $t>t'$ we go back to eq. (\ref{NSE1}) above. Observe that the left hand side of eq. (\ref{NSE1}) is just the expectation value $\left\langle f^i\left(x,t\right)v^k\left(x',t'\right)\right\rangle$. This may be contrasted to Novikov's formula \cite{Novikov,Novikov1}

\be
\left\langle f^i\left(x,t\right)v^k\left(x',t'\right)\right\rangle=\int d^dy\;N^{ij}\left(x-y\right)\left\langle \frac{\partial v^k\left(x',t'\right)}{\partial f^j\left(y,t\right)}\right\rangle,
\label{Nov}
\te
so 

\be
G^{j}_k(x,t;y,t')=\left\langle \frac{\partial v^j\left(x,t\right)}{\partial f^k\left(y,t'\right)}\right\rangle=(-i)\left\langle v^j\left(x,t\right)A_k\left(y,t'\right)\right\rangle.
\label{ge}
\te
And then, taking the Fourier transform from eq. (\ref{ge2}), we get

\be 
\left\langle \frac{\partial v^j\left(x,t\right)}{\partial f^k\left(y,t'\right)}\right\rangle=G^j_k[x-y,t,t']=
\int\frac{d^dp}{\left(2\pi\right)^d}e^{ip\left(x-y\right)}\Delta^j_k\left[p\right]G\left[p,t,t'\right].
\label{ge3}
\te
We have shown that $G$ is also related to a variational derivative of the velocity field, but in this case the derivative is taken with respect to the stirring forces, rather than the initial velocities. To complete our argument, we must show that the two derivatives are the same when the two time arguments are different. The proof depends upon the NSE being of first order with respect to time, and the stirring noise being aditive (see Appendix).

Recall that we are assuming that for given initial conditions at time $t''$, say, and a noise realization at all times between $t''$ and $t$, the solution to the NSE between $t''$ and $t$ is unique, which requires Random Galilean Invariance to be broken. Of course, this is also assumed in LET, because otherwise the function $R$ would be undefined.

Let $t'$ be an intermediate time, $t>t'>t''$. Then the solution of the NSE \ref{LIBNSE} from $t'$ to $t$ is uniquely determined by the velocity field at $t'$ and the sources for times later than $t'$, which are independent of the sources at times earlier than $t'$. Therefore, we may write

\be 
\left\langle \frac{\partial v^j\left(x,t\right)}{\partial f^k\left(y,t''\right)}\right\rangle=\int\,dy'\left\langle \frac{\partial v^j\left(x,t\right)}{\partial v^l\left(y',t'\right)}\frac{\partial v^l\left(y',t'\right)}{\partial f^k\left(y,t''\right)}\right\rangle
\te
In the limit $t'\to t^{''+}$ the  derivative with respect to the driving force may be computed from the continuum limit of eq. (\ref{DNSE}). Observe that $\partial v^l(x,t)/\partial f^k(x',t')=0$ for all times $t<t'$ because of causality. Taking the derivative of eq.  (\ref{DNSE}) with respect to the stirring force, and integrating from any negative time to $t^{''+}$, we get

\be
\frac{\partial v^l\left(y',t^{''+}\right)}{\partial f^k\left(y,t''\right)}=\Delta^l_k\left(y'-y\right),
\te 
no longer a random variable, and so we find that 

\be
\left\langle \frac{\partial v^j\left(x,t\right)}{\partial f^k\left(y,t''\right)}\right\rangle=\left\langle \frac{\partial v^j\left(x,t\right)}{\partial v^k\left(y,t''\right)}\right\rangle
\te
as we wanted to show. 

Let us mention that $G$ is discontinuos at the coincidence limit, since

\be 
\left\langle \frac{\partial v^j\left(x,t^-\right)}{\partial f^k\left(y,t\right)}\right\rangle=0;\;\;\;\left\langle \frac{\partial v^j\left(x,t^+\right)}{\partial f^k\left(y,t\right)}\right\rangle=\Delta^j_k(x-y)
\te
If we \emph{define} $G[p,t,t]=G[p,t^+,t]$, then $G=R$ everywhere.

We now turn to the discussion of the fluctuation-response relation eq. (\ref{MSRfrr}). Within the MSR approach we have the exact formula, derived from eq. (\ref{master3b}) above (assuming $t>t'$ for simplicity)

\be 
C[p,t,t']=\int^t ds\;\int^{t'}ds'\;G[p,t,s]\bar N[p,s,s']G[-p,t',s']
\label{exact}
\te 
where $\bar N$ is the dressed noise self-correlation, which comes from

\bea
\bar N^{jk}[x,t;x',t']&=&\frac{\partial^2S}{\partial\bar A_j(x,t)\partial\bar A_k(x',t')}+2\frac{\partial \Gamma_2}{\partial\left\langle A_j(x,t)A_k(x',t')\right\rangle}\nn
&=&\int\frac{d^dp}{\left(2\pi\right)^d}e^{ip\left(x-x'\right)}\Delta^{jk}\left[p\right]\bar N\left[p,t,t'\right].
\tea

For this to reduce to eq. (\ref{MSRfrr}) two conditions must be met, first

\be 
\bar N[p,s,s']=0\;{\rm{for}}\; s>t'>s'
\label{first}
\te
and second

\be  
G[p,t,s]=G[p,t,t']G[p,t',s]\;{\rm{for}}\; t>t'>s
\label{second}
\te
Since $\bar N$ is symmetric, eq. (\ref{first}) implies that $\bar N[p,s,s']=0$ for any $s\not= s'$. Thus the dressed noise must be white in time, same as the bare noise in eq. (\ref{LIBNSE}). 

With respect to the composition rule eq. (\ref{second}), the propagator $G$ obeys an  equation, also derived from eq. (\ref{master3b}) above, of the form

\be
\frac{\partial}{\partial t}G[p,t,s]+\nu p^2G[p,t,s]+\int_s^tds'\;\Sigma[p,t,s']G[p,s',s]=\delta(t-s)
\te 
where $\Sigma$ is the self-energy

\bea
\Sigma^j_k[x,t;x',t']&=&2\frac{\partial \Gamma_2}{\partial\left\langle A_j(x,t)v^k(x',t')\right\rangle}\nn
&=&\int\frac{d^dp}{\left(2\pi\right)^d}e^{ip\left(x-x'\right)}\Delta^j_k\left[p\right]\Sigma\left[p,t,t'\right].
\tea

Since we are assuming strict inequality $t>s$ the right hand side vanishes. Then eq. (\ref{second}) requires

\be 
\int_{s}^{t'}ds'\;\Sigma[p,t,s']G[p,s',s]=0,
\te 
or else 

\be 
\Sigma[p,t,s']=0\;{\rm{for}}\; t>t'>s
\te 
Once again, this means that $\Sigma[p,t,s']=0$ for any $s'\not= t$, and so the evolution of the propagator must be Markovian

\be 
\Sigma[p,t,s']=\Sigma_p(t)\delta(t-s')
\te 
in which case

\be 
G[p,t,s]=e^{-\nu p^2(t-s)-\int_s^tds'\;\Sigma_p(s')}
\te 
clearly satisfies the composition rule eq. (\ref{second}). 

In summary, \emph{if} the dressed noise is \emph{nearly} white in time and the evolution of the propagator is \emph{nearly} Markovian, then we obtain the fluctuation-response relation eq. (\ref{MSRfrr}) as a zeroth order approximation, and the proof of equivalence of the MSR and LET closures is complete. These assumptions are very strong, but not stronger than the assumption of \emph{near} Gaussianity in the equal-time velocity fluctuations in the LET approach.

\section{Final Remarks}\label{FR}

In this work we have discussed the closure problem in statistical hydrodynamic turbulence within the framework of the Martin-Siggia and Rose 2PI effective action. The formalism leads to a Bethe-Salpeter equation for the three point functions which in the leading approximation reduces to an explicit closure, reproducing the results of ref \cite{LET}. 

We believe this result validates the use of functional methods and the MSR effective action in hydrodynamic turbulence. Over and above this, it immediately suggests strategies to go beyond this leading order results, using tools such as partial resummations \cite{LP1,LP2,LP3,LP4} and the renormalization group applied to higher irreducible effective actions \cite{Meg3,Meg5,Meg6,LC17,Blaizot,MS}, which are already well developed in the literature.

\appendix 

\section{G is equal to R, an example}\label{GeR}

We would like to complement the formal arguments to the effect that the two functions $G[p,t,t']$ and $R[p,t,t']$ with a concrete example. Consider the one-dimensional noisy Burgers equation \cite{Burgers74,Bertini94,Fogedby99}

\be
\frac{\partial u}{\partial t}+\frac12\frac{\partial u^2}{\partial x}-\nu\frac{\partial^2 u}{\partial x^2}=f(x,t)
\label{noisyBurgers}
\te 
where the noise $f$ is Gaussian and white in time. We wish to compare 

\be 
R[x,t;x',t']=\frac{\delta u(x,t)}{\delta u(x',t')}
\te 
to 

\be 
G[x,t;x',t']=\frac{\delta u(x,t)}{\delta f(x',t')}.
\te
The Cole-Hopf transformation

\be 
u=-2\nu\frac{\partial }{\partial x}\ln\phi
\te 
transforms eq. (\ref{noisyBurgers}) into (assuming trivial boundary conditions in spatial infinity)

\be 
\frac{\partial\phi}{\partial t}-\nu\frac{\partial^2 \phi}{\partial x^2}=F(x,t)\phi
\label{noisyBurgers2}
\te 
where

\be 
F(x,t)=\frac{-1}{2\nu}\int_{-\infty}^xdy\;f(y,t)
\te
We may transform eq. (\ref{noisyBurgers2}) into an integral equation using the Green function

\be
H(x,t)=\frac{e^{-x^2/4\nu t}}{\sqrt{4\pi\nu t}};
\te
whereby, given $t>t'$

\be 
\phi(x,t)=\int\;dy\;H(x-y,t-t')\phi(y,t')+\int_{t'}^tds \int\;dy\;H(x-y,t-s)F(y,s)\phi(y,s).
\label{inteq}
\te
We now have

\be 
R[x,t;x',t']=\int\;dy\;\frac{\delta u(x,t)}{\delta \phi(y,t)}\frac{\delta \phi(y,t)}{\delta u(x',t')}
\label{der0}
\te 
and 

\be 
G[x,t;x',t']=\int\;dy\;\frac{\delta u(x,t)}{\delta \phi(y,t)}\frac{\delta \phi(y,t)}{\delta f(x',t')}.
\te
Now 

\bea
\frac{\delta \phi(x,t)}{\delta f(x',t')}&=&\int_{t'}^tds \int\;dy\;H(x-y,t-s)F(y,s)\frac{\delta \phi(y,s)}{\delta f(x',t')}+\int_{t'}^tds \int\;dy\;H(x-y,t-s)\frac{\delta F(y,s) }{\delta f(x',t')}\phi(y,s)\nn
&=&\int_{t'}^tds \int\;dy\;H(x-y,t-s)F(y,s)\frac{\delta \phi(y,s)}{\delta f(x',t')}-\frac1{2\nu}\int_{x'}^{\infty}\;dy\;H(x-y,t-t')\phi(y,t').
\label{der1}
\tea
On the other hand

\be 
\frac{\delta \phi(x,t)}{\delta u(x',t')}=\int\;dy\;\frac{\delta \phi(x,t)}{\delta \phi(y,t')}\frac{\delta \phi(y,t')}{\delta u(x',t')}
\label{der2}
\te 
To compute the second term, we observe that

\be 
\phi(x,t)=e^{-(1/2\nu)\int_{-\infty}^xdy\;u(y,t)},
\te 
so

\be 
\frac{\delta \phi(y,t')}{\delta u(x',t')}=\frac{(-1)}{2\nu}\phi(y,t')\theta(y-x').
\label{der3}
\te
We finally compute the first term in eq. (\ref{der2})

\be 
\frac{\delta \phi(x,t)}{\delta \phi(y,t')}=H(x-y,t-t')+\int_{t'}^tds \int\;dy'\;H(x-y',t-s)F(y',s)\frac{\delta \phi(y',s)}{\delta \phi(y,t')}.
\label{der4}
\te
Using eqs. (\ref{der3}) and (\ref{der4}) in (\ref{der2}) we obtain

\be 
\frac{\delta \phi(x,t)}{\delta u(x',t')}=\int_{t'}^tds \int\;dy\;H(x-y,t-s)F(y,s)\frac{\delta \phi(y,s)}{\delta u(x',t')}-\frac1{2\nu}\int_{x'}^{\infty}\;dy\;H(x-y,t-t')\phi(y,t').
\label{der5}
\te

This establishes the identity of $R$ and $G$, since they share both the same equation and the same initial data.

\acknowledgements

I thank the anonymous referee to Manuscript FW10261 (\cite{C25}) who suggested this project.

E. C. acknowledges financial support from Universidad de Buenos Aires through Grant No. UBACYT
20020170100129BA, CONICET Grant No. PIP2017/19:11220170100817CO and ANPCyT Grant No. PICT 2018: 03684.


\begin{thebibliography}{999}



\bibitem{EF10} G. Eyink and U. Frisch, Robert H. Kraichnan, in P. Davidson et al. (editors) \emph{A Voyage through Turbulence} (Cambridge U.P., Cambridge, 2011).


\bibitem{Wyld61}{Wyld, H.W., Jr. Formulation of the Theory of Turbulence in an
Incompressible Fluid. \emph{Ann. Phys.} \textbf{1961}, \emph{14}, 143--165.}

\bibitem{Lee65} L. L. Lee, A Formulation of the Theory of Isotropic Hydromagnetic
Turbulence in an Incompressible Fluid, Ann. Phys. 32, 292 (1965).

\bibitem{LB87} V. I. Belinicher and V.S. L'vov, A scale-invariant theory of fully developed hydrodynamic turbulence, Zh. Eksp. Teor. fiz. 93, 533 (1987) (Engl. Trans. Sov. Phys. JETP 66, 303 (1987)).


\bibitem{MSR73} P.C. Martin, E.D. Siggia and H.A. Rose, Statistical Dynamics of Classical Systems, Phys.
Rev. A8 423 (1973).

\bibitem{Ph77} R Phythian, The functional formalism of classical statistical dynamics, J. Phys. A: Math. Gen., Vol. 10, No. 5, 777 (1977).

\bibitem{Eyink} G.L. Eyink, Turbulence Noise, J. Stat. Phys. 83, 955  (1996).

\bibitem{Kamenev} A. Kamenev, \emph{Field Theory of Non-Equilibrium Systems}, Cambridge University Press,
Cambridge, U.K. (2011).

\bibitem{JZEC} J. Zanella and E. Calzetta, Renormalization group and nonequilibrium action in stochastic
field theory, Phys. Rev. E 66 036134 (2002).

\bibitem{DeDGia06} C. de Dominicis and I. Giardina, \emph{Random fields and spin glasses: a field theory approach} (Cambridge University Press, Cambridge, England, 2006).


\bibitem{Krommes1978} J. A. Krommes, Turbulence, clumps and the Bethe-Salpeter equation (IAEA-SMR-32/10), in \emph{Theoretical and Computational Plasma
Physics}, International Atomic Energy Agency, Vienna, p. 405 (1978).

\bibitem{Krommes1997} J. A. Krommes, Systematic statistical theories of plasma turbulence
and interrnittency: current status and future prospects, Physics Reports 283, 5 (1997).

\bibitem{Krommes2002} J. A. Krommes, Fundamental statistical descriptions of plasma turbulence in magnetic  fields, Physics Reports 360, 1 (2002).




\bibitem{BS2} Bethe, H.A. and Salpeter, E. E., A Relativistic Equation for Bound-State Problems, Phys. Rev. 84, 1232 (1951).

\bibitem{IZ} C. Itzykson and J-B. Zuber, \emph{Quantum field Theory}, McGraw-Hill, New York (1980).

\bibitem{TK} L. Tsang and J. Kong, \emph{Scattering of electromagnetic waves: Advanced topics}, John Wiley, New York (2001).

\bibitem{Mariano} M. Franco and E. Calzetta, Wave propagation in non-Gaussian random
media, J. Phys. A: Math. Theor. 48 (2015) 045206.


\bibitem{LET} W D McComb and S R Yoffe, A formal derivation of the local energy transfer
(LET) theory of homogeneous turbulence, J. Phys. A: Math. Theor. 50,  375501 (2017).


\bibitem{vKH}Th. de Karman and L. Howarth, On the statistical theory of
isotropic turbulence, Proc. R. Soc. A 164, 192 (1938).

\bibitem{Karman48} T.  von Kármán, Progress in the statistical theory of turbulence. Proc. Natl.
Acad. Sci. USA 34, 530 (1948). 

\bibitem{Chandra} S. Chandrasekhar, \emph{The theory of turbulence}, editado por E. Spiegel, Springer (2011).

\bibitem{MonYag71} A. S. Monin and A. M. Yaglom, \emph{Statistical Fluid Mechanics}, MIT Press, 1971.

\bibitem{FRIS95}  U. Frisch, \emph{Turbulence, the Legacy of A. N. Kolmogorov} (Cambridge University Press, Cambridge, England, 1995).


\bibitem{Novikov} E. A. Novikov, Random force method in turbulence theory, J. Exptl. Theoret. Phys. (U.S.S.R.) 44, 2159-2168 (1963) (Eng. trans. Sov. Phys. JETP 17, 1449 (1965)).

\bibitem{Novikov1} E. A. Novikov, Functionals and the random-force method in turbulence theory, J. Exptl. Theoret. Phys. (U.S.S.R.) 47, 1919-1926 (1964) (Eng. trans. Sov. Phys. JETP 20, 1290 (1965)).


\bibitem{CalHu08} E. Calzetta and B-L. Hu, \emph{Nonequilibrium Quantum Field Theory} (Cambridge University Press, Cambridge, England, 2008).



\bibitem{Ram07} J. Rammer, \emph{Quantum field theory of nonequilibrium states} (Cambridge University Press, Cambridge (England), 2007)

\bibitem{Haehl18}{F. M. Haehl, R. Loganayagam and M. Rangamani, Effective action for relativistic hydrodynamics: fluctuations, dissipation, and entropy inflow, JHEP 10, 194 (2018).}

\bibitem{MGKC} N. Mirón Granese, A. Kandus and E.
Calzetta,  Field Theory Approaches to Relativistic
Hydrodynamics. Entropy  24,
1790 (2022).

\bibitem{C25} E. Calzetta, Energy spectrum of non-Newtonian turbulence, Phys. Rev. Fluids 10, 054607 (2025).

\bibitem{LutWar60} J. Luttinger and J. Ward, Ground-state energy of many-fermion systems, Phys. Rev. \textbf{118}, 1417 (1960).

\bibitem{DomMar64b} C. de Dominicis and P. C. Martin, Stationary entropy principle and
renormalization in normal and superfluid systems II, J. Math. Phys.
\textbf{5}, 31 (1964).

\bibitem{DahLas62} H.D. Dahmen and G. Jona Lasino, Variational Formulation of Quantum Field Theory, Il Nuovo Cimento LIIA, 807 (1967).

\bibitem{cornwall:1974a} J.M. Cornwall, R. Jackiw and E. Tomboulis, Effective action for composite operators, Phys. Rev. D {\bf 10}, 2428 (1974).



\bibitem{Meg4} M. E. Carrington, B. A. Meggison, and D. Pickering, 2PI effective action at four loop order in $\varphi^4$ theory, Phys. Rev. D 94, 025018 (2016)

\bibitem{Gabor} M.E. Carrington, G. Kunstatter, H. Zaraket, 2PI effective action and gauge dependence identities, Eur. Phys. J. C 42, 253–259 (2005).

\bibitem{Emil} E. Mottola, Gauge invariance in 2PI effective actions, Strong and Electroweak Matter 2002, pp. 432-436 (2003)

\bibitem{Urko} U. Reinosa, Renormalization and gauge symmetry for 2PI effective actions, in Strong and Electroweak Matter 2004, 316, World Scientific (2005) arXiv:hep-ph/0411255v1


\bibitem{C09} E. Calzetta, Kadanoff-Baym equations for near-Kolmogorov turbulence, arXiv:0908.4068.


\bibitem{Bode} Tim Bode, The two-particle irreducible effective action
for classical stochastic processes, J. Phys. A: Math. Theor. 55 (2022) 265401.


\bibitem{dch} E. Calzetta and B. L. Hu, Decoherence of correlation histories, in \emph{Directions in General Relativity,
Vol II: Brill Festschrift}, edited by B. L. Hu and T. A. Jacobson
~Cambridge University Press, Cambridge, England, 1993.

\bibitem{cddn} E. Calzetta and B. L. Hu, Correlations, Decoherence, Dissipation, and Noise in Quantum Field Theory, in \emph{Heat Kernel Techniques and
Quantum Gravity}, edited by S. A. Fulling, Texas A\&M Press,
College Station, TX, 1995, hep-th/9501040.

\bibitem{stobol} E. Calzetta and B. L. Hu, Stochastic dynamics of correlations in quantum field theory:
From the Schwinger-Dyson to Boltzmann-Langevin equation, Phys. Rev. D, VOLUME 61, 025012 (1999).

\bibitem{stobol2} E. Calzetta, Fourth-order full quantum correlations from a
Langevin–Schwinger–Dyson equation, J. Phys. A: Math. Theor. 42 (2009) 265401.

\bibitem{Berges} Jürgen Berges, n-particle irreducible effective action techniques for gauge theories, Phys. Rev. D, VOLUME 70, 105010 (2004).

\bibitem{Meg1} M. E. Carrington, WeiJie Fu, T. Fugleberg, D. Pickering, and I. Russell, Bethe-Salpeter equations from the 4PI effective action, Phys. Rev. D 88, 085024 (2013).

\bibitem{Meg2} M.E. Carrington, Techniques for calculations with nPI effective actions, EPJ Web of Conferences ,04013 (2015).


\bibitem{Hopf} E. Hopf, Statistical hydromechanics and functional calculus, J. Rat. Mech. Anal. 1, 87-123 (1952).


\bibitem{Zin93}  J. Zinn-Justin, \emph{Quantum Field Theory and Critical Phenomena} (Clarendon Press, Oxford, 1993).


\bibitem{LP1} Victor L’vov and Itamar Procaccia, Exact resummations in the theory of hydrodynamic turbulence. I. The ball of locality and normal scaling, Phys. Rev. E52, 3840 (1995).

\bibitem{LP2} Victor L’vov and Itamar Procaccia, Exact resummations in the theory of hydrodynamic turbulence. II. A ladder to anomalous scaling, Phys. Rev. E52, 3858 (1995).

\bibitem{LP3} Victor L’vov and Itamar Procaccia, Exact resummations in the theory of hydrodynamic turbulence. III. Scenarios for anomalous scaling and intermittency, Phys. Rev. E53, 3468 (1995).

\bibitem{LP4} Victor L’vov and Itamar Procaccia, Towards a nonperturbative theory of hydrodynamic turbulence: Fusion rules, exact bridge
relations, and anomalous viscous scaling functions, Phys. Rev. E54, 6268 (1996).




\bibitem{Meg3} M. E. Carrington, Wei-Jie Fu, D. Pickering, and J.W. Pulver, Renormalization group methods and the 2PI effective action, Phys. Rev. D 91, 025003 (2015).

\bibitem{Meg5} M. E. Carrington, S. A. Friesen, B. A. Meggison, C. D. Phillips, D. Pickering, and K. Sohrabi, 2PI effective theory at next-to-leading order using the functional renormalization group, Phys. Rev. D 97, 036005 (2018).

\bibitem{Meg6} Margaret E. Carrington and Christopher D. Phillips, Four Loop Scalar $\phi^4$ Theory Using the Functional
Renormalization Group, Universe 2019, 5, 9.

\bibitem{LC17} Federico Lamagna and Esteban Calzetta, A functional renormalization method for
wave propagation in random media, J. Phys. A: Math. Theor. 50 (2017) 315102.

\bibitem{Blaizot} Jean-Paul Blaizot, Jan M. Pawlowski, Urko Reinosa, Functional renormalization group and 2PI effective action formalism, Annals of Physics 431 (2021) 168549.


\bibitem{MS} Peter Millington and Paul M Saffin, Vertex functions and their flow equations
from the 2PI effective action, J. Phys. A: Math. Theor. 55 (2022) 435402.

\bibitem{Burgers74} J. Burgers, \emph{The nonlinear diffusion equation}, Springer (1974).

\bibitem{Bertini94} L. Bertini, N. Cancrini, G. Jona-Lasinio, The Stochastic Burgers Equation, Commun. Math. Phys. 165, 211-232 (1994).  

\bibitem{Fogedby99} H. Fogedby, Aspects of the noisy Burgers equation, in \emph{Anomalous Diffusion From Basics to Applications}, Lecture Notes in Physics, Volume 519. Springer-Verlag (1999), p. 101.




\end{thebibliography}
\end{document}